# Acoustic-to-Articulatory Speech Inversion Features for Mispronunciation Detection of /ɹ/ in Child Speech Sound Disorders

*Nina R Benway\**[1], *Yashish M Siriwardena\**[2], *Jonathan L Preston*[1,3], *Elaine Hitchcock*[4], *Tara McAllister*[5], *Carol Espy-Wilson*[2]

[1]Communication Sciences and Disorders, Syracuse University, New York, USA
[2]Electrical and Computer Engineering, University of Maryland College Park, Maryland, USA
[3]Haskins Laboratories, New Haven, Connecticut, USA
[4]Communication Sciences and Disorders, Montclair State University, New Jersey, USA
[5]Communicative Sciences and Disorders, New York University, New York, USA
\*denotes equal contribution; `nrbenway@syr.edu, yashish@umd.edu`

## Abstract

Acoustic-to-articulatory speech inversion could enhance automated clinical mispronunciation detection to provide detailed articulatory feedback unattainable by formant-based mispronunciation detection algorithms; however, it is unclear the extent to which a speech inversion system trained on adult speech performs in the context of (1) child and (2) clinical speech. In the absence of an articulatory dataset in children with rhotic speech sound disorders, we show that classifiers trained on tract variables from acoustic-to-articulatory speech inversion meet or exceed the performance of state-of-the-art features when predicting clinician judgment of rhoticity.

## 1. Introduction

Our recent work has effectively automated evidence-based treatment for speech sound disorders [1]. While best-practice clinician-led motor-based interventions include both summary perceptual judgment feedback (i.e., *knowledge of results*; KR) and detailed auditory/somatosensory corrective feedback (i.e., *knowledge of performance*; KP) [2], no available automated treatment delivers clinical-grade KP. To date, automated systems deliver KR with clinician-delivered KP [3] or approximated KP using random selection of plausible clinical cues (often mimicking clinician-delivered KP due to obscured visualization of the tongue in the oral cavity, [4]). However, a mispronunciation detection system that leverages acoustic-to-articulatory speech inversion (SI) may someday automate true corrective KP that is not attainable with state-of-the-art formant-based systems, while also circumventing known issues in (child) formant estimation [5].

**Acoustics and Articulation of American English /ɹ/**

### 1.1.1. Articulatory Configurations for /ɹ/

We focus on rhotic /ɹ/, the most commonly impacted sound in American English speech sound disorders persisting past age 8 [6]. For fully rhotic /ɹ/, there is evidence for up to five quasi-dependent articulatory actions in the oral and pharyngeal cavities: (1) elevation of the tongue tip/blade, (2) lateral bracing of the tongue against the posterior molar teeth, (3) a low posterior tongue dorsum, (4) retraction of the tongue root into the pharynx, and (5) slight rounding of the lips [4]. Insufficient vocal tract configurations will likely generate a more neutral spectral envelope that may be perceived as a "derhotic" /ɹ/. Derhotic vocal tract configurations commonly include a lower tongue tip, insufficient tongue root retraction, and, notably, a higher tongue dorsum [4, 7]. Indeed, blade/dorsum relative displacement has been found to be salient for predicting perception of the syllable /ɑɹ/ in speech sound disorders [8].

### 1.1.2. Acoustic Configuration of /ɹ/

Although there is variation in the vocal tract configurations that generate a perceptually-correct American English /ɹ/, these configurations generate a similar spectral envelope in a linear predictive coding (LPC) formant feature space. Formant features serve as the baseline condition for the present work because they are well-motivated by decades of acoustic phonetics investigation for /ɹ/ and also meet reproducibility guidelines advocating low-dimension, validated feature sets in clinical speech technology systems [9]. Furthermore, our recent work has demonstrated that age-and-sex normalized formants outperform cepstral representations in the binary classification of /ɹ/ rhoticity in the context of speech sound disorders [10]. Prior acoustic phonetics investigations show that vocal tract configuration for a correct, fully rhotic /ɹ/ is marked by a relatively high second formant (F2) [11] and a relatively low third formant (F3) [12]. This results in the average rhotic F3-F2 distance being much narrower than that of a neutral vocal tract. LPC estimation, however, requires speaker customization and can be error-prone [13, 14], particularly for samples taken from (child) speakers with high fundamental frequencies [5]. Errors in estimation may be speaker-specific [15] or due to population-general traits such as wider bandwidths in children [14]. Our experience in manually correcting formant estimates in the context of /ɹ/ also indicates that near merging of F3 and F2 in hyper-rhotic tokens is an additional source of estimation error.

**Motivation and Contributions**

Speech analysis systems that estimate the learner's articulatory trajectory have the potential to improve performance of state-of-the-art formant-based rhoticity classifiers. Such systems could eventually automate the delivery of KP feedback at or above the level possible by human clinicians, due to the low visual salience of the vocal tract during for /ɹ/. Because a ground truth articulatory dataset is not available for children with rhotic speech sound disorders, de novo training of SI for this use case is not possible at this time. Therefore, in this study we test the performance of an adult-trained SI system on utterance-normalized child speech data to improve over a formant baseline for the binary classification of /ɹ/ rhoticity in children with speech sound disorders.

We offer two contributions. Our first research question shows that tract variable estimates generated from adult-trained models are able to index clinician perceptual judgment of rhotic errors, particularly for tongue body constriction location *(d = .39)*. Our second research question shows that a bidirectional LSTM trained on tract variable output from SI meets or exceeds the performance of a comparable system trained on age-and-gender normalized formants when predicting clinician judgment of rhoticity ($\bar{x}_{F1\text{-}score}$ = .90 $\sigma_x$ = .05; med = .92, n = 6).

## 2. Speech Inversion

Acoustic-to-articulatory SI is tasked with retrieving articulatory dynamics from a speech signal [16]. Attempts at recovering articulatory movements from the continuous speech signal have a long history [17], but have generally limited to tracking a specific set of articulators like upper and lower lip, tongue tip, velum closure, etc. However, it is important to derive not just the main effect of individual articulator movements, but the interactions among articulators; for example, lips and jaw as individual articulators work together to achieve a desired vocal tract shape [18]. Hence, general SI systems focus on recovering the vocal tract constriction, estimating the constriction degree and location of functional tract variables (TVs), rather than the movement of individual articulators. During SI, acoustic features extracted from the speech signal are used to predict the TVs. The inverse mapping is learned by associating these features through training on a corpus consisting of matched acoustic and directly observed articulatory data.

Here, we use the SI system developed in [19], which is a Temporal-Convolution Network (TCN) trained on 36 (adult) speakers from the U. Wisconsin X-Ray microbeam (XRMB) corpus [20], and evaluated with no speaker overlap in training, development, and test splits. Apart from the 6 TVs (LA: Lip Aperture, LP: Lip Protrusion, TTCL: Tongue Tip Constriction Location, TTCD: Tongue Tip Constriction Degree, TBCL: Tongue Body Constriction Location, TBCD: Tongue Body Constriction Degree), the system predicts 3 source features: aperiodicity, periodicity, and pitch.

## 3. Methods

This study is a binary classification experiment seeking to predict listener judgment of rhoticity (i.e., fully rhotic/ "correct" vs derhotic/ "incorrect") in a subset of the open access PERCEPT-R audio Corpus [21] collected during a clinical trial of biofeedback interventions [22]. The PERCEPT subset selected for reanalysis are 3,210 word-level utterances from the 6 speakers providing consent/assent for future use of study audio (publicly available at [23]; see [24] for additional corpus audio and ground-truth label details which include multi-listener average perceptual ratings of rhoticity). Original data collection received ethics approval from the Biomedical Research Alliance of New York. Speech data were lab-collected by research clinicians during word-reading probes using head-mounted microphones. Although a small sample, these data were chosen for this exploration for several reasons. Firstly, they are clinically-valid group of children with speech sound disorder. Secondly, the range of speaker ages in this dataset (Table 1) allows for preliminary exploration of the impact of child age on model performance. Thirdly, because these data come from a treatment study in which some participants experienced statistically and clinically significant gains in /ɹ/ production, these data contain tongue shape variation between pretreatment to posttreatment timepoints within the same speaker. Lastly, a subset of the audio has hand-placed, within-word segmentation boundaries for the rhotic target.

Table 1: *Participants in the current investigation*

| PERCEPT-R ID | Study ID | Age | Formant Ceiling | Number of Utterances |
|---|---|---|---|---|
| 33 | 6102 | 15.7 | 6000 Hz | 543 |
| 34 | 6103 | 14.9 | 4500 Hz | 638 |
| 35 | 6104 | 9.3 | 6000 Hz | 560 |
| 36 | 6108 | 14.5 | 5000 Hz | 692 |
| 37 | 3101 | 9.8 | 5500 Hz | 337 |
| 38 | 3102 | 11.8 | 5000 Hz | 440 |

**Formant Extraction**

Time series estimates of F1, F2, and F3 were extracted from the utterance using custom Python scripts and the Praat "To Formant (Robust)" command with default settings except for participant-specific Formant Ceiling settings. Formant ceilings were customized using the procedure in [22]. Formant transform time series (F3-F2 distance and F3-F2 deltas were also included in the feature set. Age-and-sex norming was completed as in [10] using a third-party dataset [25].

**Tract Variable Extraction**

We extract 6 TVs from the SI system to capture the degree and location of the lip, tongue body and tongue tip constrictions. The extracted TVs (in the range of [-1,1]) are z-normalized, utterance-wise, to generate the final 6 TV feature set. We additionally estimate glottal activity by extracting three source level features (aperiodicity, periodicity and pitch) [26]. Similar to the 6 TVs, the 3 source features are also z-normalized, utterance-wise. The 6 TVs and 3 source features together comprise the 9 TV feature set.

**Rhotic Segment Boundary Estimation**

Research assistants who were trained in speech signal segmentation manually reviewed each pre- and post-treatment utterance to mark the onset and offset of the rhotic phoneme using Praat TextGrids. Only pretreatment and posttreatment files were selected for this analysis because data collection at these timepoints elicited citation speech, rather than in-treatment speech that may be marked by unnaturally long or unstable articulations. Within each utterance, research assistants selected the segment corresponding to the perception of the (target) rhotic phoneme. Coarticulatory transitions between target rhotics and neighboring segments were wholly assigned following the sonority hierarchy, so transitions were included with the more sonorous segment. In other words, rhotic-vowel transitions were included with the vowel, while rhotic-consonant transitions were included with the rhotic. Boundary decisions were made based on visual breakpoints in F2 slope and confirmed perceptually during file playback. If F2 breakpoints were not discernable, F3 was used, and then F1. We used these rhotic boundary timestamps to extract all rhotic-associated TVs, which we then averaged into 10 bins to facilitate visual inspection for our first research question (only).

**Prediction of Clinician Perceptual Judgment using Leave One Participant Out Cross Validation**

The differentiation of derhotic and rhotic segments by formants and tract variables was assessed with Deep Neural Network (DNN) based models, which can be extremely effective in binary classification tasks. We specifically experimented with Recurrent Neural Networks (RNN) due to the timeseries nature of the data. The source code for all model implementations will be made publicly available upon publication of the paper.

*3.1.1. Ground Truth Labels*

The binary class label for the audio files present investigation was derived from the listener-average binary perceptual rating in the PERCEPT-R Corpus (i.e., 1 = fully rhotic/"correct" vs 0 = derhotic/"incorrect"). For files with non-unanimous listener ratings, .66 served as the floor for class 1 (the fully rhotic class) to reflect the lack of full agreement between expert raters in the context of RSSD [27]. All utterances with a listener-average rating < .66 were assigned to class 0 (the derhotic class).

*3.1.2. Data Preprocessing*

The five age-and-sex normalized formants and transforms described in 3.1 and the TVs described in 3.2 were used independently as input features for model training. The features were segmented or padded to generate 2-second-long input embeddings. Each input embedding was matched to a ground-truth label for rhoticity from the PERCEPT-R Corpus, representing the average binary listener rating (0 = derhotic, 1 = fully rhotic). The heuristic for discretizing the average rating to binary ground-truth in the present investigation was .66.

*3.1.3. Model Architecture and Training*

We experimented with Bidirectional gated RNN (BiGRNN) and Bidirectional LSTM (BiLSTM) models with different architectures. Figure 2 shows the best performing BiLSTM model architecture for the classifier.

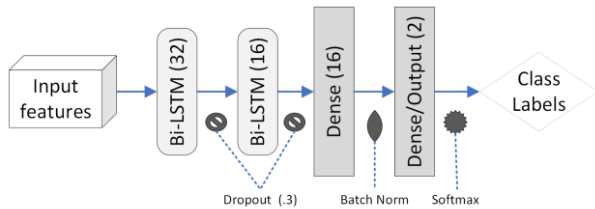

Figure 2: *Best-performing BiLSTM model architecture*

All model parameters were randomly initialized with a seed (=7) for reproducibility of the results. All the models were trained using leave-one-participant-out cross validation. Models in each training split were trained with features from ~2300 utterances (~85% of non-test data). The validation set was chosen at random and included ~15% of non-test data. Models were trained for 50 epochs with an early stopping criterion and a patience of 5 epochs. Since the dataset is imbalanced (favoring derhotic samples, as would be expected in speech sound disorder clinical trial data), a weighted cross entropy loss function was used to optimize the models.

*3.1.4. Hyperparameter Fine-Tuning*

Table 2 lists candidate hyperparameters and corresponding values for the best performing BiLSTM model. Hyper-parameters were fine-tuned with grid search optimizing validation loss. All models were implemented with the keras/TensorFlow framework and trained with NVIDIA TITAN X GPUs. The best performing model included ~123k trainable parameters, and converges in ~ 7 min. for each validation split.

Table 2. *Tuned hyperparameters (bold) with candidate values*

| Parameter | Candidate Values |
|---|---|
| **Learning Rate** | [**1e-4**, 3e-4, 1e-3, 1e-2] |
| **Batch size** | [16, 32, **64**, 128] |
| **Optimizer** | **ADAM**, RMSprop, SGD |
| **BiGRNN/BiLSTM Layers** | [1, **2**, 3, 4] |
| **Dense Layers** | [1, **2**, 3] |
| **Dropout** | [**.3**, .5, .7] |

# 4. Results

**Effect Size Calculation for Tract Variables**

Our first research question analyzed the univariate ability of (time-binned) tract variables to distinguish between child speech productions with a ground truth label corresponding to "derhotic" and "fully rhotic". We quantified the separation of the derhotic and fully rhotic tract variable means using Cohen's d. Mean separation was "negligible" for lip aperture ($d = -.06$), lip protrusion ($d = .11$), tongue tip constriction location ($d = .13$), and tongue tip constriction degree ($d = .08$). Mean separation was "small" for tongue body constriction location ($d = .39$; 95%CI [.33, .44]; ) and for tongue body constriction degree ($d = -.22$; 95%CI [-.27, -.16]), as seen in Figure 3. The lower values for tongue body constriction location imply a more anterior constriction in fully rhotic tokens than derhotic tokens.

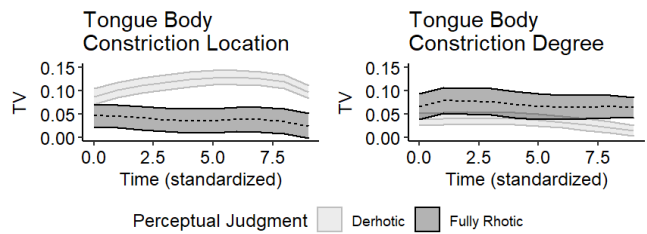

Figure 3. *Univariate differentiation of perceptual judgment for (binned) TVs. Ribbons: 95% confidence intervals of the mean.*

**Predicting clinician judgment of rhoticity in utterances from children with RSSD**

Our second research question analyzed the performance of age-and-sex normalized formant and tract variable feature sets. Results from BiGRNN and BiLSTM models are shown in Table 3. BiLSTM improved participant-specific F1-score (weighted) over BiGRNN. BiLSTM performance was comparable for formant and 9 TV feature sets, except for the notable increase in AUROC in the context of 9 TV features (Figure 4).

Table 3: *Mean (standard deviation) of participant-specific performance. 9 TVs include 3 source features.*

| Model | Feature Set | F1-Score | Precision | Recall | AUROC |
|---|---|---|---|---|---|
| **Bi-GRNN** | Formants | .83 (.05) | .90 (.05) | .79 (.06) | .82 (.05) |
|  | 6 TVs | .72 (.10) | .89 (.05) | .67 (.10) | .76 (.06) |
|  | 9 TVs | .81 (.06) | .91 (.04) | .77 (.07) | .80 (.04) |
|  | Formants | .89 (.03) | .92 (.04) | **.88 (.03)** | .79 (.17) |
| Bi-LSTM | 6 TVs | .82 (.08) | .91 (.04) | .77 (.11) | .81 (.13) |
| | 9 TVs | **.90 (.05)** | **.94 (.04)** | **.87 (.08)** | **.87 (.07)** |

The interactions between participants and feature sets (Figure 4) naturally raised the question of combined performance, so we trained one Bi-LSTM with a combined feature set of formants and 9 TVs. Performance ($\bar{x}_{F1\text{-}score}$ = .82 $\sigma_x$ = .05) was lower than in models trained on these features individually.

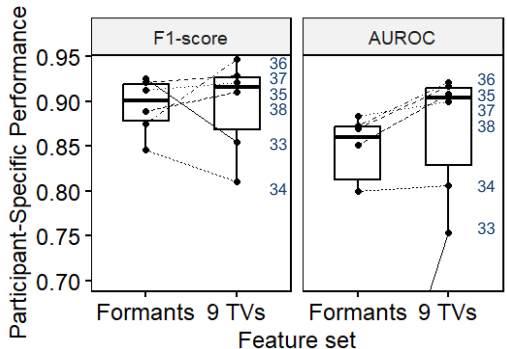

Figure 4. *Bi-LSTM performance for individual participants (labels on the right). Not shown: formant AUROC for 33 (.44).*

Although the number of participants in this reanalysis was limited, we explored the correlation of ranks to see if participant age was associated with AUROC. We selected AUROC for exploration as its participant-specific scores showed greater variance than F1-score in our results. Spearman's rho was not significant in this sample ($\rho$ = -.54; p = .30, n = 6); this is supported by visual inspection of Figure 5 which illustrates that children aged 9-14 achieved AUROC > .9. Notably, the lowest performing participants in this sample had vocal tracts that, presumably, shared the most age-based similarities with the adult participants that the SI system was originally trained on.

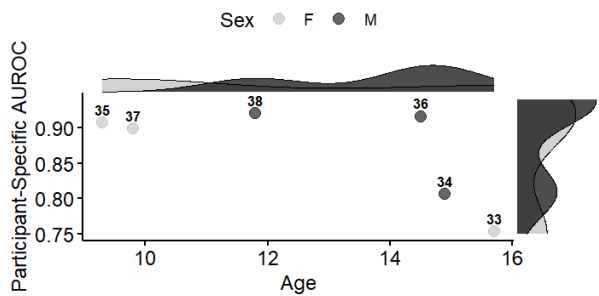

Figure 5. *Participant-specific Bi-LSTM performance (AUROC) by participant age and sex. Labels = PERCEPT IDs*

## 5. Discussion

The present classification study predicts binary listener perceptual judgment of rhoticity in child speakers with speech sound disorder impacting /ɹ/. We found that BiLSTM performance was largely comparable when trained on (1) hand-crafted age-and-sex normalized formant features and (2) tract variable + source feature (9 TV) outputs generated from an adult-trained acoustic-to-articulatory SI system. In this sample, however, AUROC was notably higher for 9 TVs than formant features. The comparison of classifier performance in 6 TV and 9 TV feature sets evidences the importance of modeling source features as additional input, consistent with observations of TV performance in depression [28]. Notably, univariate separation between classes for individual TVs indicates that tongue dorsum dynamics may be an important signal for clinical /ɹ/, which is corroborated by magnetic resonance/ultrasound imaging of American English /ɹ/ perceived as clinically incorrect [4, 7] and supportive of recent modeling work elsewhere [8]. Future reanalysis of ultrasound data in [22] could investigate if this finding might associate with direct instruction of a retroflexed (versus bunched) configuration.

Even though this TV estimator was trained on adult speech, there was no evidence in the present, small, sample that TV performance was systematically related to age. However, it was surprising that the oldest participants in this sample had the (relatively) poorest performance. Informal observations during [22] indicate that participant 34's improvements in /ɹ/ resulted in a consistently hyper-rhotic articulatory pattern with a merged F3-F2 by treatment session 3, and participant 33 had a large F3-F2 distance not due to a high F3, but because of a low F2. Future investigation can quantify the extent that poorer performance may be due to speaker-specific vocal tract dynamics.

Although this investigation is limited by a small sample size, it motivates larger investigation of how adult-trained TV estimates perform for child speech. This motivation arises from the 1) lack of obvious age effects for younger speakers using 9 TVs, 2) advantage for 9 TVs in AUROC, and 3) lower amount of participant-specific customization required by TVs in feature preprocessing. Larger samples can also investigate model explainability, as well as evaluate if lower performance herein on the fused formant + 9TV feature set is evidence of true performance or perhaps due to the high feature dimensionality relative to the amount of training data for the 6 participants.

Lastly, the salience of TVs in this investigation lay the foundation for future research modeling interpretable KP feedback for speech sound learners from the 9 TVs. In addition to predicting perceptual judgment of /ɹ/, visualizations of tongue shape interpolated from TVs have the potential to provide similar benefits to ultrasound biofeedback while circumventing some of the barriers associated with widespread ultrasound use (i.e., system cost and training).

## 6. Conclusions

Overall, BiLSTM models trained individually on hand-crafted age-and-sex normalized formants and 9 tract variables (predicted from an adult-trained acoustic-to-articulatory speech inversion system) performed similarly when predicting clinical perceptual judgment of rhoticity in child speech sound disorders. However, improvements in AUROC and lower customization needs for tract variable generation, as well as the potential for clinically interpretable KP feedback, motivate future development of tract variable-based mispronunciation detection for /ɹ/. This work further motivates the collection of ground-truth articulatory data from children to validate tract variables for child clinical speech sound technologies.

## 7. Acknowledgements

Funding was provided by National Institute on Deafness and Other Communication Disorders (NIH R01DC017476-S2, T. McAllister, PI), and the National Science Foundation (IIS1764010, S. Shamma & C. Espy-Wilson, PIs), and was supported in part through computational resources provided by Syracuse University (NSF ACI-1341006; NSF ACI-1541396).